\documentclass[onecolumn,showpacs,superscriptaddress,showkeys,preprint,prb]{revtex4-1}
\usepackage{helvet}
\usepackage{graphicx}
\usepackage[latin1]{inputenc}
\usepackage{color}
\begin{document}

\author{Marten Richter}
\email[]{marten.richter@tu-berlin.de}

\affiliation{Institut für Theoretische Physik, Nichtlineare Optik und
Quantenelektronik, Technische Universität Berlin, Hardenbergstr. 36, 10623
Berlin, Germany}
\affiliation{Department of Chemistry, University of California, Irvine,
California 92697-2025, USA}

\author{Felix Schlosser}
\author{Mario Schoth}
\affiliation{Institut für Theoretische Physik, Nichtlineare Optik und
Quantenelektronik, Technische Universität Berlin, Hardenbergstr. 36, 10623
Berlin, Germany}

\author{Sven Burger}
\author{Frank Schmidt}
\affiliation{Zuse Institute Berlin, Takustr. 7, 14195 Berlin, Germany}
\author{Andreas Knorr}
\affiliation{Institut für Theoretische Physik, Nichtlineare Optik und
Quantenelektronik, Technische Universität Berlin, Hardenbergstr. 36, 10623
Berlin, Germany}

\author{Shaul Mukamel}
\affiliation{Department of Chemistry, University of California, Irvine,
California 92697-2025, USA}

\title{Reconstruction of the wavefunctions of coupled nanoscopic emitters using a coherent optical technique}

\begin{abstract}
{  We show how} coherent, spatially resolved spectroscopy can disentangle complex hybrid wave functions 
  into  wave functions of the individual  emitters.
This way, detailed information on the coupling of the individual emitters, not available in far-field spectroscopy, {  can be revealed}.
Here we propose a quantum state tomography { protocol}
that relies on the ability to selectively excite each emitter individually by 
spatially localized  pulses. 
Simulations of coupled semiconductor GaAs/InAs quantum dots using light fields available in current nanoplasmonics show, that   undesired resonances can be removed from measured spectra. 
The method can be applied on a broad range of coupled emitters to study the internal coupling, including pigments in photosynthesis and artificial light harvesting.
 \end{abstract}

\date{August 10, 2012}

\maketitle
\section{Introduction}
The formation of collective optical resonances from Coulomb-coupled  optical emitters  is a very general phenomenon,
including examples from chromophores  in biological light harvesting complexes\cite{Engel:Nature:07,Christensson:JPhysChemLett:10,Solvay::11,Schoth:PhysRevLett:12}, semiconductor quantum dots \cite{Guenther:PhysRevLett:02,Lovett:PhysRevB:03}, metal nanoparticles and composite systems, such as plasmon lasers \cite{Bergman:PhysRevLett:03}.

For all  these structures, dipole-dipole coupling occurs on a nanometer scale and the  states of the individual emitters   hybridize to form new collective, so called  excitonic states, delocalized over the whole structure.
Far field excitation, governed by the wavelength resolution limit $\lambda/2$, can only probe delocalized exciton states of a nanostructure. Related far-field experiments such as absorption, pump probe and four wave mixing  are unable to disentangle the individual contributions of the coupled emitters from the collective optical response, because the exciting fields are spatially constant on the scale of the entire structure and  cannot discriminate different emitters. In contrast, spatially  local spectroscopy such as near field spectroscopy can, in principle, address   the individual emitters.

In this paper, we propose a new 
class of measurements  that combine coherent nonlinear spectroscopy with near field optics
to reconstruct { the contributions of single emitters to the delocalized wave function }in a spatially extended   nanostructure.
As an example,
we demonstrate,  how a coherent double-quantum-coherence optical technique \cite{Abramavicius:ChemRev:09} may be
 combined with spatially localized fields   to reconstruct the { exciton} wave functions of three dipole coupled self-organized GaAs/InAs quantum dots. This constitutes a particular quantum state tomography. { The presented procedure is independent of the  technique for localizing the fields at  individual emitters. Several localization methods are known and are already applied to a broad range of nanoemitting structures, e.g. using (metalized) near field fiber tips \cite{vonFreymann:ApplPhysLett:98,Guenther:ApplPhysLett:99,Guenther:PhysRevLett:02},
  metal tips \cite{Pettinger:PhysRevLett:04,Weber-Bargioni:NanoLett:11}, nano antennas \cite{Zhang:NanoLett:09,Kinkhabwala:NatPhoton:09,Schuller:NatMat:10,Curto:Science:10,Novotny:PhyscisToday:11} and metal structures combined with pulse shaped fields \cite{Stockman:PhysRevLett:02,Aeschlimann:Nature:07}.}

Quantum state tomography  is a development aimed at the direct reconstruction of  wave functions or more generally the density matrix, first proposed  
by Fano \cite{Fano:RevModPhys:57}.
  The importance of  quantum state tomography results from the fact, that the reconstruction and knowledge of the wave function opens the possibility to calculate new observables not related to optics at all. Examples include magnetic moments and transport properties. 
So far, wave functions are seldom directly accessible by experiments \cite{Gerhardt:PhysRevA:10}.
Recent advances include imaging 
of single orbitals  using  soft-x-ray pulses 
\cite{Kapteyn:Science:07,Corkum:NatPhys:07} and the  reconstruction of  states \cite{Yuen-Zhou:JChemPhys:11,Lobino:Science:08}. 
 Applications  so far range from Spin 1/2 particles \cite{Band:AmJPhys:79},  photon states  using the Wigner function \cite{Vogel:PhysRevA:89,Smithey:PhysRevLett:93}, vibrational states \cite{Dunn:PhysRevLett:95} to Josephson junctions \cite{Steffen:Science:06}.  
{In contrast to earlier approaches, the
quantum state tomography developed in this paper combines} optical fields, highly localized in time and space 
with coherent  2D spectroscopy,   using  a sequence of light pulses with controlled envelopes and phases \cite{Abramavicius:ChemRev:09,Li:PhysRevLett:06,Aeschlimann:Science:11}.

\section{Excitons in coupled nanostructures}
\begin{figure}[tb]
\center
a)\includegraphics[width=2.25 cm]{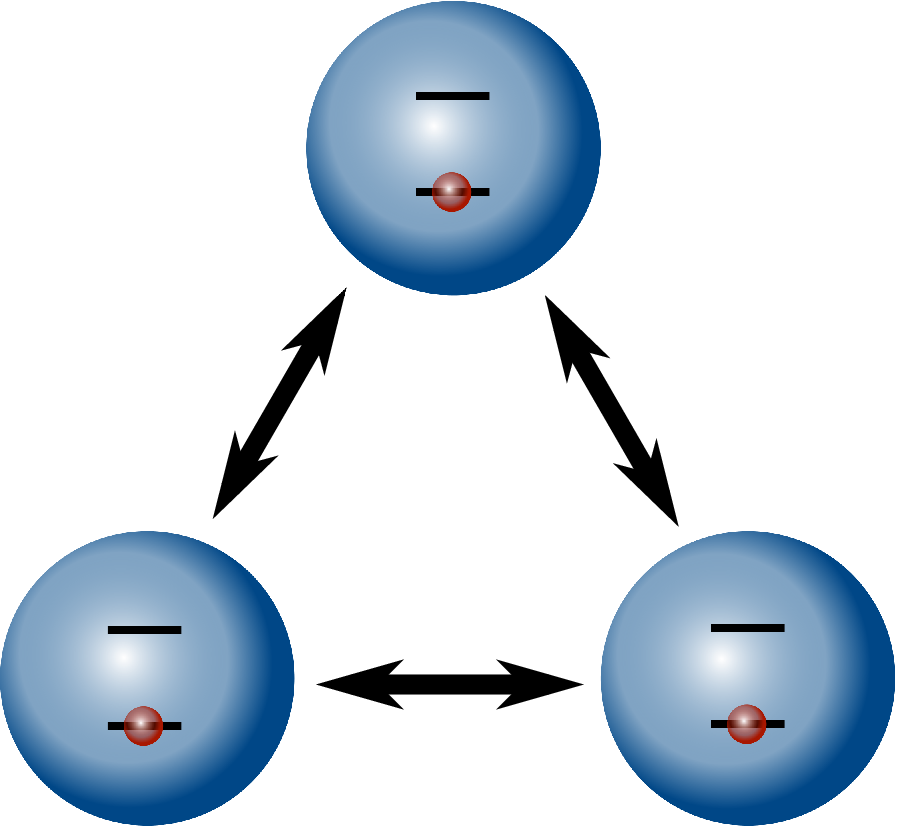}
b)\includegraphics[width=1.5 cm]{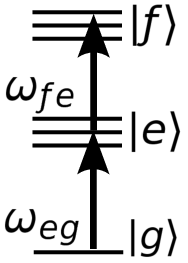}
\caption{  a) Three dipole-dipole coupled self-organized InAs quantum dots,
b) Exciton level scheme of the three coupled quantum dots.} 
\label{schemes_cpl_qds}
\end{figure}

As a typical example for coupled nanostructures with delocalized wave functions, we study
 three coupled self-organized semiconductor quantum dots \cite{Lovett:PhysRevB:03,Danckwerts:PhysRevB:06,Dachner:PhysStatusSolidiB:10}, cp. Fig.~\ref{schemes_cpl_qds}a). 
The quantum dot distance is  assumed to be sufficiently large to have no electronic wave function overlap 
between the quantum dots. 
In this case we study interdot coupling in the form of dipole-dipole (or Förster) coupling  known from
 selforganized GaAs/InAs quantum dots: Parameters like dot size, dot distances,  coupling constants, and energy shifts
 are well known from theory \cite{Richter:PhysStatusSolidiB:06} and experiment \cite{Unold:PhysRevLett:05}.
 { Each quantum dot is represented as a two level system.
This is a valid assumption for quantum dots provided
 (i) quantum dots have no spin-orbit splitting and a big enough biexcitonic shift,
 (ii) are negatively charged or (iii)  have spin-orbit coupling bigger 
 than the inter quantum dot couplings \cite{Jacak::98,Borri:PhysRevLett:01}.
 }
For selforganized  quantum dots with sizes of $20 \mathrm{nm}$ and interdot distances around $40 \mathrm{nm}$,
the dipole coupling is about  several $\mu \mathrm{eV}$  with a Lorentzian zero phonon line (ZPL) width 
of $\gamma=1 \mu \mathrm{eV}$  {  at low temperatures (e.g. $T=4 K$) \cite{Borri:PhysRevLett:01,Stock:PhysRevB:11}}.
{ We neglected the influence of 
the phonon side bands, since their amplitude in the spectra is
one to two orders smaller than the amplitude of the zero phonon line resonance at 
low temperatures \cite{Borri:PhysRevLett:01,Stock:PhysRevB:11}. 
}

Three coupled quantum dots exhibit joint states: a ground state  $g$,
three single-exciton  states 
  $e_1$, $e_2$ and $e_3$  and  three two-exciton  states $f_1$, $f_2$ and $f_3$, cf. Fig.~\ref{schemes_cpl_qds}b).
  The system has one triexciton state, but these states are of no relevance in a third order optical experiment, considered here.
The ground state of the uncoupled quantum dots  is not changed by the induced dipole-dipole coupling.
The delocalized single-exciton states $|e\rangle$ resulting from the dipole-dipole interaction
 are composed of local, uncoupled quantum dot states  $|i\rangle$ (quantum dot  $i$ in excited state):
 $|e\rangle =
\sum_{i} c^e_i 
|i\rangle$. $|e\rangle$ is an
energy eigenstate of the coupled quantum dot system, $c^e_i$ the expansion coefficients. 
Similarly, two-exciton states $|f\rangle$  are
composed of states with two local excitations at quantum dot i
and j: $|f\rangle =\sum_{i< j} c^f_{ij}|ij\rangle$.
In general, excited states of
\protect$N$  coupled two level system
emitters  
form  a ground state \protect$g$, \protect$N$ delocalized single-exciton states \protect$e$ and \protect$N(N-1)/2$ delocalized  two exciton states \protect$f$.
For our three dot case, we choose  couplings between two quantum dots  slightly stronger than to the third quantum dot
({parameters} given in Table \ref{hamop}). Here, $H_0$ includes along the diagonal the transition frequency local emitters modified by single and two exciton shifts, respectively. The offdiagonal elements describe interactions describing excitation transfer caused by e.g. dipole-dipole interactions.

\begin{table}[b]
\center
a)
\begin{tabular}{|l|lll|}
\hline
$\langle i|H_0|j\rangle$ & $1$ & $2$& $3$  \\
\hline
$1$ & $2.0$ & $1.0$ & $0.2$ \\
$2$ & $1.0$ & $0.2$ & $0.1$ \\
$3$ & $0.2$ & $0.1$ & $-2.5$\\
\hline
\end{tabular}\\
b)
\begin{tabular}{|l|lll|}
\hline
$\langle i j|H_0|kl\rangle$ & $1,2$ & $1,3$& $2,3$  \\
\hline
$1,2$ & $\langle 1|H_0|1\rangle+\langle 2|H_0|2\rangle+V_{t1}$ & $\langle 3|H_0|2\rangle$ & $\langle 3|H_0|1\rangle$ \\
$1,3$ &$\langle 2|H_0|3\rangle$ & $\langle 1|H_0|1\rangle+\langle 3|H_0|3\rangle+V_{t2}$& $\langle 2|H_0|1\rangle$ \\
$2,3$ & $\langle 1|H_0|3\rangle$ & $\langle 1|H_0|2\rangle$ & $\langle 1|H_0|1\rangle+\langle 3|H_0|3\rangle+V_{t3}$\\
\hline
\end{tabular}
\caption{Hamiltonoperator in matrix form. a) The single exciton block and b) the two exciton states block. All values are given  in $\mu eV$. The diagonal elements of the matrices are given as detuning to a mean gap frequency, a) $\omega_{gap}$ for the single excitons, b) $2\omega_{gap}$ for the two excitons, with $\omega_{gap}=1.053\mathrm{eV}$ and the two exciton shifts $V_{t1}=0.1 \mu eV$, $V_{t2}=-2.5 \mu eV$ and $V_{t3}=-1.5 \mu eV$.}
\label{hamop}
\end{table}

First, to characterize the system within far field spectroscopy, we calculate
the linear absorption spectrum:
\begin{eqnarray}
 \alpha(\omega)\propto\sum_{e} \frac{|\mu_{eg}|^2}{(\omega-\omega_{eg})^2+\gamma^2} .
\end{eqnarray}
Here, $\mu_{eg}$ is the dipole moment for ground state to single-exciton transition, $\omega_{eg}$ is the transition frequency and $\gamma$ the dephasing constant.

\begin{figure}[tb]
\center
\includegraphics[width=5 cm]{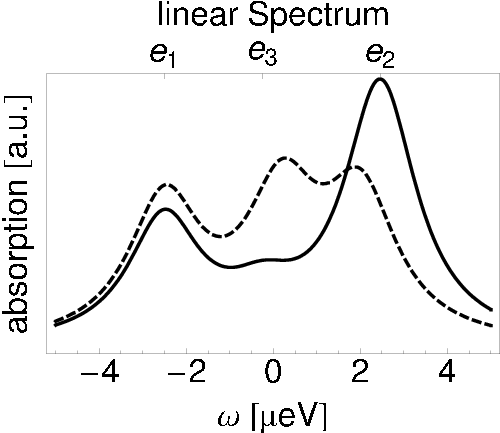} 
\caption{  Absorption spectrum coupled (solid) and uncoupled (dashed) quantum dots. The detection frequency $\omega$ is given as detuning relative to frequency  $\omega_{gap}=1.053\mathrm{eV}$
(transition frequency of uncoupled quantum dot 3).} 
\label{linearspec}
\end{figure}

The absorption spectrum of the coupled quantum dot structure is plotted in Fig. \ref{linearspec}(solid). The single-exciton states $e_1$, $e_2$, $e_3$  overlap spectrally  such that only  $e_1$ and $e_2$ are well resolved,  $e_3$  contributes only with a spectral shoulder. Comparing coupled and uncoupled(dashed) spectra, one recognizes, that the oscillator strength is originally evenly distributed but strongly modified, since the dipole-dipole coupling forms excitons delocalized over the entire structure. 

\section{Ingredients for reconstructing delocalized states}
{\it Our main goal is to gain information on the built up of the delocalized wavefunctions of the excitonic states,
i.e. on the expansion coefficients $c^e_i$, for a given
single-exciton state~$|e\rangle$. }
For this purpose, we use  coherent, spatially local spectroscopy, composed of three ingredients:\\
(i) local nanoscale excitation provided by metallic nanoantennas and  refined pulse shaping techniques \cite{Aeschlimann:Nature:07,Reichelt:OptLett:09} to optically address individual quantum dots, (Section \ref{localized_excitation})\\
(ii) phase cycling of the optical response \cite{Meyer:ApplPhysB:00,Tian:Science:03,Brinks:Nature:10},  to disentangle the total nonlinear response into desired quantum paths, (Section \ref{phase_cycling_sec}) \\
(iii) a postprocessing procedure  to calculate the coefficients $c^e_i$ (Section \ref{extract_wavefunc}).

{\it  In general, (ii) and (iii) can be applied to any quantum system
representable by spatial separated coupled emitters, 
if any localization technique (i) is available.}

\subsection{Localized excitation} \label{localized_excitation}
\begin{figure}
\center
a)\includegraphics[width=3.8cm]{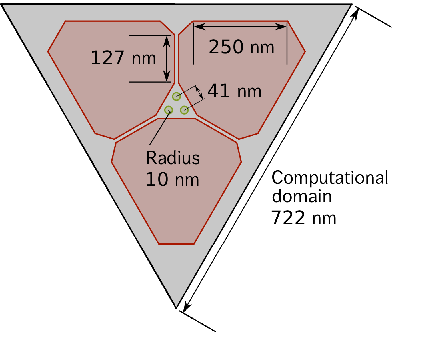} 
 b)\includegraphics[width=5 cm]{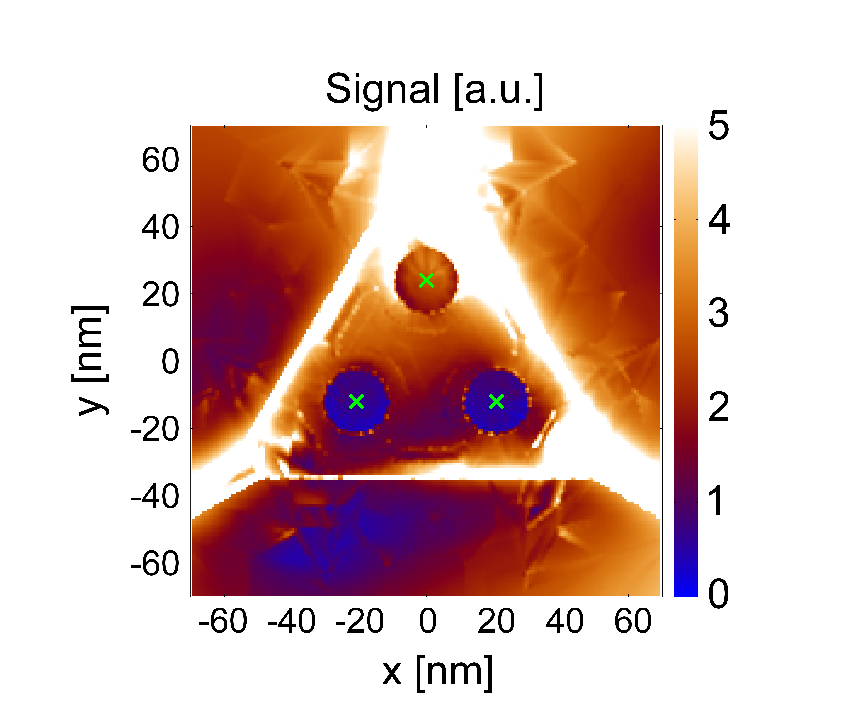} 
  \caption{a) Schematic geometry: Three  selforganized GaAs/InAs quantum dots (diameter $20 \mathrm{nm}$, inter dot distances $40 \mathrm{nm}$) and three $12\,\mathrm{nm}$ thick silver layer structures, arranged with 120° rotational symmetry on a  semi infinite GaAs-layer.
   b)  Optimized localized electric field $|E|$   at the maximum peak  for a single pulse composed from shaped pulses from three different directions.
Note: the white color is $5$ arbitary units or higher.
The magnitude (extracted  at X) of the optimized electric field $|E(\mathbf{r},\omega_{gap})|$ for $\omega_{gap}$ at dots 2 and 3 is $12\%$ or $7.5\%$ compared to dot 1.
{ The field intensity is scalable while  third order perturbation theory  is valid for  the double quantum coherence spectrum.}
}
\label{geometry}
\end{figure}

A main ingredient of our scheme is the local excitation of individual quantum dots.
In our specific example, we achieve local excitation of the individual quantum dots 
by a plasmonic antenna structure of triangular symmetry on a subwavelength scale, cp. Fig. \ref{geometry}a). 
These metal structures can be realized by e-beam lithography.
Solving Maxwell's equations for this geometry shows that  plasmonic effects and an  optimization procedure of the applied pulses allows to selectively excite single quantum dots \cite{Brixner:PhysRevLett:05,Pomplun:physstatussolidib:07}:

For optimizing the pulse envelope of a single pulse $\mathbf{E}(t,\mathbf{r})$ { towards a field localization at only one quantum dot},
we use time-harmonic solutions $\mathbf{E}_\nu(\omega, \mathbf{r})$,  represented by incident plane waves of polarization directions $\mathrm{p},\mathrm{s}$ and   incoming direction (indexed as $\nu$)\cite{Pomplun:physstatussolidib:07}:
\begin{eqnarray}
\mathbf{E}(t,\mathbf{r}) = \frac{1}{\sqrt{2\pi}}\int_{-\infty}^{\infty} d\omega \sum_\nu g_\nu(\omega)\mathbf{E}_\nu(\omega, \mathbf{r})e^{-\imath \omega t}.
\end{eqnarray}
Pulse shaping is introduced by  the weighting function:
\begin{eqnarray}
g_\nu(\omega) &=& \sum_{n} f^\nu(\vartheta_n) \frac{A_n}{\sqrt{2\pi}}e^{-(\eta_n-\omega)^2{\sigma_n}^ {2}/{2}+\imath\omega\tau_n+\imath\beta_n} \label{param_genalg},
\end{eqnarray}
which represents a composition of Gaussian pulses with amplitudes $A$, center times $\tau$,  frequencies $\eta$, widths $\sigma$, phases $\beta$, and polarization angle $\vartheta$ for each pulse $n$~projected to polarization direction $\nu$ ($f^\mathrm{p} = \cos, f^\mathrm{s} = \sin$).
$g_{\nu}(\omega)$ has to be determined by optimization.
To increase the number of optimization parameters, we combine the three incoming pulses from three directions, using  $120°$ symmetry of the sample.
For this paper, details of the optimization procedure are of no relevance but 
can be found in Ref. \onlinecite{Brixner:PhysRevLett:05,Reichelt:OptLett:09}.  Later on, the absolutes value of $E(t)$ in the quantum dots centers  is the input for the calculation of the localized  spectra.

In Fig. \ref{geometry}b), the spatial field distribution for  the optimized total field around the quantum dot transition
frequency is shown.
It can be recognized, that
a  chosen, single quantum dot is excited stronger than the other quantum dots.
We observe field enhancements
between different quantum dot sites of a factor of eight or larger.
{ Note, that the optimized fields in  frequency domain  
show that polarization and propagation phase effects cause localization 
and not a frequency based selection of different quantum dots.}

{
Note, that the presented localization scheme using  excitation pads is just an example. 
For application of the protocol to other systems  \cite{Guenther:PhysRevLett:02,vonFreymann:ApplPhysLett:98,Guenther:ApplPhysLett:99,Pettinger:PhysRevLett:04,Weber-Bargioni:NanoLett:11,Zhang:NanoLett:09,Kinkhabwala:NatPhoton:09,Schuller:NatMat:10,Curto:Science:10,Novotny:PhyscisToday:11,Stockman:PhysRevLett:02,Aeschlimann:Nature:07}, other spatial localization schemes might be used.
  }

\begin{figure}[tbh]
\center
\includegraphics[width=8 cm]{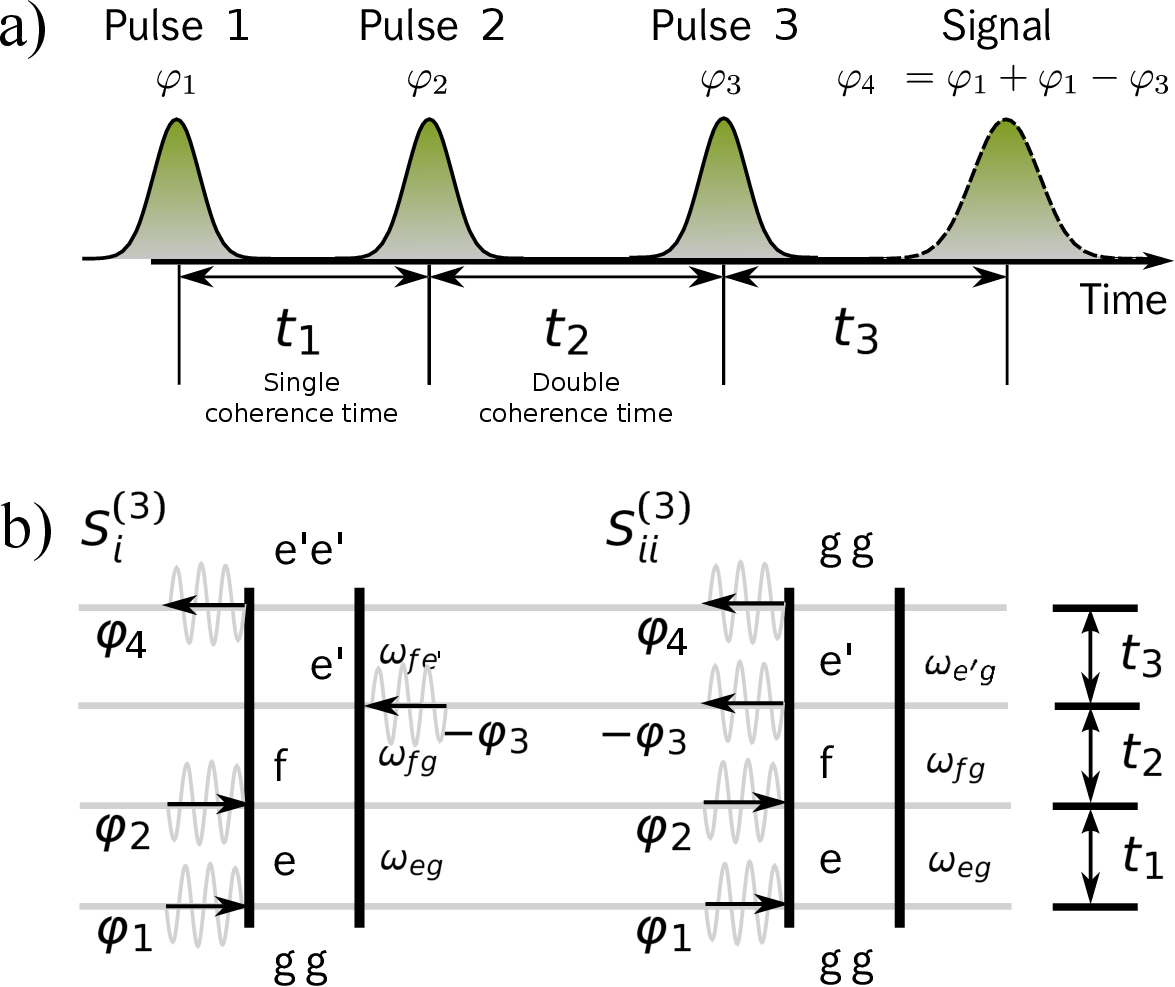}
\caption{ 
 a) Pulse sequence  for the   double quantum coherence experiment.
 b) The two density matrix pathways
$S_i$ and $S_{ii}$ ( Eq. \ref{k3locfirst}). See text for more details.
} 
\label{lioupathways}
\label{pulseconfiguration}
\end{figure}

\begin{figure}[tbh]
\center
\includegraphics[width=8.3 cm]{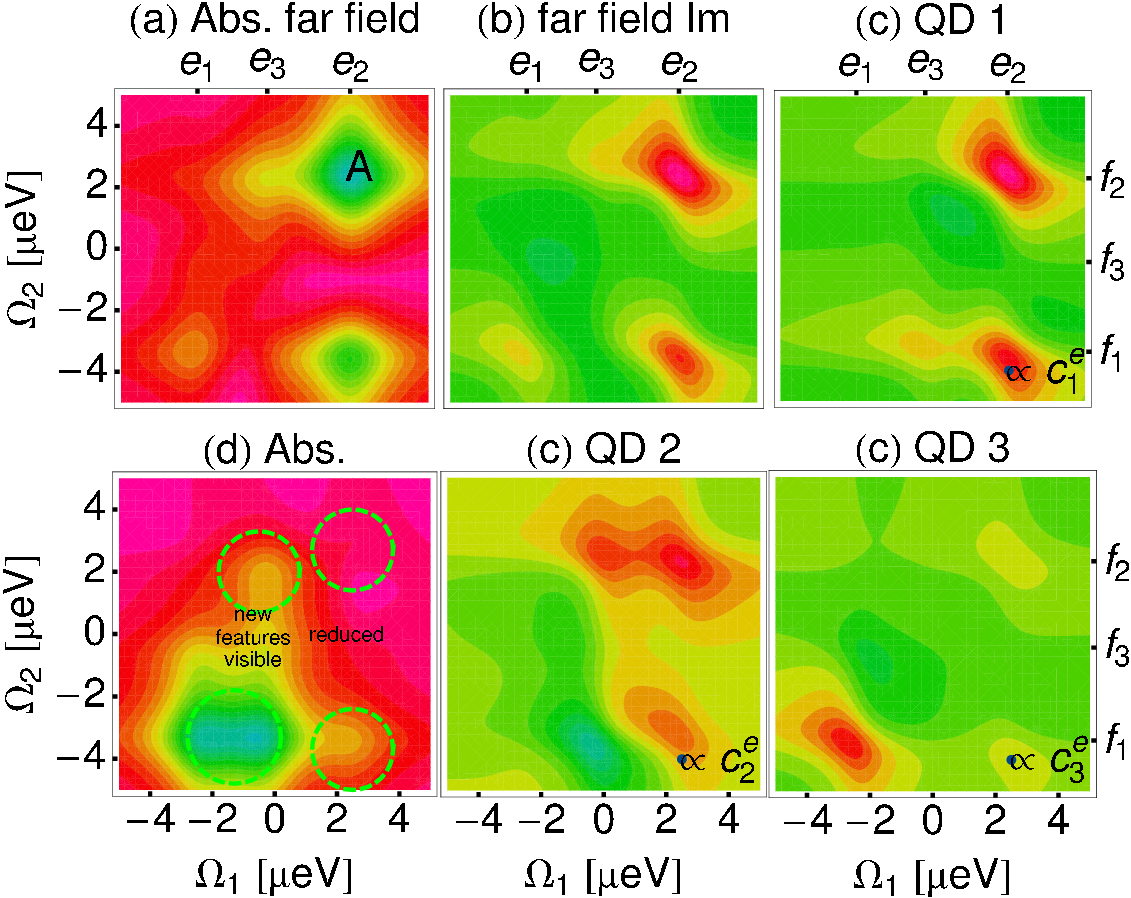}

\caption{ 
 DQS for $t_3=200\mathrm{ps}$ a) absolute value  b) imaginary part, $\Omega_1$ ($\Omega_2$) given as detuning around  the single (double) gap frequency $\omega_{gap}$.
 c) Imaginary part of localized double quantum coherence spectrum, where the first pulse is  localized at quantum dot 1, 2 and 3 as indicated d) Filtered standard double quantum coherence spectrum ($e_2$ removed).  
 }

\label{SDQClocimag}
\label{standardDQC}
\end{figure}

\subsection{Phase cycling detection of coherent signals }\label{phase_cycling_sec} 
As explained in Sec. \ref{localized_excitation},
a  sequence of three spatially optimized pulse envelopes $E^i$ with phases $\varphi_i$ and laser frequency $\omega_l$  is used \cite{Yang:PhysRevLett:08,Abramavicius:ChemRev:09}:
\begin{eqnarray}
E(\mathbf{r},t)&=&E^1(\mathbf{r},t-t_3-t_2-t_1)e^{\imath \omega_l (t-t_3-t_2-t_1)+\imath \varphi_1}\nonumber\\
&&
+E^2(\mathbf{r},t-t_3-t_2)e^{\imath \omega_l (t-t_3-t_2)+\imath \varphi_2}\nonumber\\
&&+E^3(\mathbf{r},t-t_3)e^{\imath \omega_l (t-t_3)+\imath \varphi_3}+c.c..\label{pulses}
\end{eqnarray}
Here the envelopes $E^i(\mathbf{r},t)$ are determined by the optimization procedure for localized pulses.
The detected signal (selected quantum pathways of the full dipole density)   is measured with heterodyne detection via phase cycling \cite{Meyer:ApplPhysB:00,Kato:ChemPhysLett:01,Tian:Science:03,Brinks:Nature:10}
by repeating
the experiment  several times for different phases $\varphi_1$, $\varphi_2$ and $\varphi_3$, cf. Fig. \ref{pulseconfiguration}a).

In general the polarisation, created by three pulses applied to the quantum dots,
is described by many quantum pathways in Liouville space \cite{Abramavicius:ChemRev:09}.
In the following way, we can extract a subset of the Liouville pathways by extracting a certain phase combination of
$\varphi_1$, $\varphi_2$ and $\varphi_3$:
{ The detected dipole density for different phases can be written as\cite{Meyer:ApplPhysB:00}: 
\begin{eqnarray}
&&P(t,\varphi_1,\varphi_2,\varphi_3)=\mathcal{P}(t,\varphi_1,\varphi_2,\varphi_3)+c.c.,\nonumber\\
&&\mathcal{P}(t,\varphi_1,\varphi_2,\varphi_3)=\sum_{lmn} c_{123,lmn} P_{lmn}(t),
\end{eqnarray}
with $c_{123,lmn}=e^{\imath (l\varphi_1+m\varphi_2+n\varphi_3)}$, $l+m+n=1$  and $|l|+|m|+|n|=1$ or $3$ for resonant excitation and $P_{lmn}(t)$ being the part of the detected polarisation with phase dependence $l\varphi_1+m\varphi_2+n\varphi_3$.
$c_{123,lmn}$ can be viewed as a matrix with first index $(\varphi_1, \varphi_2, \varphi_3)$   
and second index $(l,m,n)$.
Carring out the experiment for sufficient phase combinations $\varphi_1$, $\varphi_2$, $\varphi_3$, so that the matrix $c_{123,lmn}$ is invertable, we can extract the signal with a specific phase combination $\varphi_4=l\varphi_1+m\varphi_2+n\varphi_3$ (selecting particular pathways) using:  $ P_{lmn}(t)=\sum_{1,2,3} {c^{-1}}_{123,lmn} \mathcal{P}(t,\varphi_1,\varphi_2,\varphi_3)$. Details of this phase cycling procedure can be found in Ref. \onlinecite{Meyer:ApplPhysB:00}}.
Typical examples for such signals are the photon-echo $\varphi_4= -\varphi_1+\varphi_2+\varphi_3$, anti-photon-echo $\varphi_4= \varphi_1-\varphi_2+\varphi_3$ (cf. Ref. \onlinecite{Abramavicius:ChemRev:09}). 

\subsection{Double quantum coherence signal} \label{doublequantumcoherencesignal}
We focus on  the double quantum coherence signal, a third order signal with  the contributing  phase combinations $\varphi_4=\varphi_1+\varphi_2-\varphi_3$   \cite{Yang:PhysRevLett:08,Abramavicius:ChemRev:09}.
In the case of a system, where the ground state, single exciton and two exciton states form three bands (cf. Fig. \ref{schemes_cpl_qds} b)), only the two Liouville pathways depicted in Fig. \ref{lioupathways}b), will contribute to the signal with 
$\varphi_4=\varphi_1+\varphi_2-\varphi_3$   \cite{Yang:PhysRevLett:08,Abramavicius:ChemRev:09}.

In the case of the three band model (Fig. \ref{schemes_cpl_qds}b)) only two Liouville pathways can contribute.
The part of the polarization attributed to $\varphi_4$, i.e. $ P^{(3)}_{1,1-1}(t)$
which depends on the delay times can be written using a reponse function  \cite{Abramavicius:ChemRev:09}:
\begin{eqnarray}
&& P^{(3)}_{1,1-1}(t)=\nonumber\\
&&\qquad\int_0^\infty \mathrm{d}\tau_3 \int_0^\infty \mathrm{d}\tau_2 \int_0^\infty \mathrm{d}\tau_1
R^{(3)}_{1,1-1}(t,t-\tau_3,t-\tau_3-\tau_2,t-\tau_3-\tau_2-\tau_1).
\end{eqnarray}
Note, that we include the optical fields into the definition of the response which is rather uncommon, 
but  for the use of localized fields this notation will simplify the discussion.
The response function $R^{(3)}_{1,1-1}$ can be divided into the contributions of  two Liouville pathways,  extracted from the full response function \cite{Abramavicius:ChemRev:09}:
\begin{eqnarray}
 &&R^{(3)}(t,\tilde{t}_3,\tilde{t}_2,\tilde{t}_1)=\left(\frac\imath\hbar\right)^3 \mathrm{tr}(\mathbf{\mu}\mathcal{G}(t-\tilde{t}_3)H_{el-L,-}(\tilde{t}_3)\mathcal{G}(\tilde{t}_3-\tilde{t}_2)\nonumber\\
 && \qquad H_{el-L,-}(\tilde{t}_2)\mathcal{G}(\tilde{t}_2-\tilde{t}_1)H_{el-L,-}(\tilde{t}_1)\rho_0). \label{response}
\end{eqnarray}
Here, the electron electric field interaction Liouvillian $H_{el-L,-}(t)\rho=[H_{el-L}(t),\rho]$, the Green function $\mathcal{G}(t)$ with $\mathcal{G}(t)\rho(t)=\theta(t)\mathrm{exp}(-\frac\imath\hbar H_0 t)\rho(t) \mathrm{exp}(\frac\imath\hbar H_0 t)$
and the dipole operator $\mu=\sum_i \mu_{gi} |g\rangle \langle i|_l+h.a.$.
For our excitonic  three band system,  for the far field excitation we insert the light matter Hamiltonian in local basis: 
\begin{eqnarray}
H_{el-L}=\sum_i \mu_{gi}\cdot E(t) |g
 \protect\rangle \protect\langle
 i|+ \sum_{ij} \mu_{gi}\cdot E(t) |j
  \protect\rangle \protect\langle
 ij| +H.a. . \label{licht_matter}
\end{eqnarray}
The Hamilton operator can also be reformulated in the delocalized basis:
\begin{eqnarray}
 H_{el-L}=\sum_e \mu_{ge}\cdot E(t) |g \protect\rangle \protect\langle e|+\sum_{ef} 
 \mu_{ef}\cdot E(t) |e \protect\rangle  \protect\langle f| +H.a. ,\label{licht_matter_delocalized}
 \end{eqnarray}
 with the delocalized exciton dipole matrix elements $\mu_{ge}=\sum_i c^{e}_i \mu_{gi}$ and  $\mu_{ef}=\sum_{i<j} {c^{e}_i}^* \mu_{gi}c^{f}_{ij}$.
We insert  Eq. (\ref{licht_matter_delocalized}) into Eq. (\ref{response}) and collect
for $R^{(3)}_{1,1-1}$  only the terms proportional to $\mathrm{exp}[\imath (-\varphi_3+\varphi_2+\varphi_1)]$
and end up with  the response from two contributing  Liouville pathways (Fig. \ref{lioupathways}b)) assuming no temporal pulse overlap \cite{Abramavicius:ChemRev:09}:
\begin{eqnarray}
&&R^{(3)}_{1,1-1}(t,\tilde{t}_3,\tilde{t}_2,\tilde{t}_1)=R^{(3)}_{i}(t,\tilde{t}_3,\tilde{t}_2,\tilde{t}_1)+R^{(3)}_{ii}(t,\tilde{t}_3,\tilde{t}_2,\tilde{t}_1)\\
&&R^{(3)}_{i}(t,\tilde{t}_3,\tilde{t}_2,\tilde{t}_1)=- \left(\frac\imath\hbar\right)^3 
e^{\imath \omega_l (\tilde{t}_1+\tilde{t}_2-\tilde{t}_3-t_1-2t_2-t_3)} \nonumber\\
&&\qquad \sum_{ee'f} \mu_{e'f} \mu_{ge'}\cdot E^{3*}(\tilde{t}_3-t_3) \mu_{fe}\cdot E^2(\tilde{t}_2-t_3-t_2)
\mu_{eg}\cdot E^1(\tilde{t}_1-t_3-t_2-t_1) \nonumber\\ 
&& \qquad e^{-\imath \xi_{fe'}(t-\tilde{t}_3)-\imath \xi_{fg}(\tilde{t}_3-\tilde{t}_2)-\imath \xi_{eg} (\tilde{t}_2-\tilde{t}_1)}\\
&&R^{(3)}_{ii}(t,\tilde{t}_3,\tilde{t}_2,\tilde{t}_1)= \left(\frac\imath\hbar\right)^3 
e^{\imath \omega_l (\tilde{t}_1+\tilde{t}_2-\tilde{t}_3-t_1-2t_2-t_3)} \nonumber\\
&&\qquad \sum_{ee'f} \mu_{ge'} \mu_{e'f}\cdot E^{3*}(\tilde{t}_3-t_3) \mu_{fe}\cdot E^2(\tilde{t}_2-t_3-t_2)
\mu_{eg}\cdot E^1(\tilde{t}_1-t_3-t_2-t_1)\nonumber\\  
&&\qquad e^{-\imath \xi_{e'g}(t-\tilde{t}_3) -\imath \xi_{fg}(\tilde{t}_3-\tilde{t}_2)-\imath \xi_{eg} (\tilde{t}_2-\tilde{t}_1)}
\end{eqnarray}
Here, $\xi_{nm}=\omega_{nm}-\imath \gamma_{nm}$, with $\omega_{nm}=\omega_n-\omega_m$ including the exciton frequencies $\omega_n$ and the dephasing/relaxation rate $\gamma_{nm}$ for a Lorentzian dephasing model.

In both pathways (i,ii), we have    a coherence between the single exciton and ground state in between the first and second pulse and   a two exciton to ground state coherence in between the second and third pulse.
After the third pulse the system is either in a single exciton to two-exciton coherence (pathway (i)) or ground state to single-exciton coherence (pathway (ii)). 
We consider for further analysis the heterodyne detected signal, where the emitted signal $P^{(3)}_{1,1,-1}(t)$ is mixed with the field of a local oscillator $E_4$:
\begin{eqnarray}
 S^{(3)}_{k_{III}}(t_1,t_2,t_2)=\int_{-\infty}^{\infty} \mathrm{d}t P^{(3)}_{1,1-1}(t) E^{4*}(t)e^{\imath \omega_l t}
\end{eqnarray}
 $S^{(3)}_{k_{III}}(t_1,t_2,t_2)$ is a complex quantity. A measurement obtains the  real part of  $S^{(3)}_{k_{III}}(t_1,t_2,t_2)$ \cite{Abramavicius:ChemRev:09}.
However the use of a local oscillator in heterodyne detection allows -by twisting its phase - to detect also the imaginary part of the signal \cite{Abramavicius:ChemRev:09,Brixner:JChemPhys:04,Zhang:ProcNatlAcadSci:07,Li:PhysRevLett:06,Christensson:JPhysChemLett:10,Dai:PhysRevLett:12} (phase cycled detection of fluorescence in fourth order\cite{Aeschlimann:Science:11} can give similar information as heterodyne detected signals in  third order), this works both for the signal in temporal and Fourier domain.
It is therefore a prefered method to extract also the phase information of the coefficients $c^e_i$, most  other methods will only allow to extract the absolute value. 

\noindent In order to separate the different coherences of the signal by their energies,
the signal is Fourier transformed over the delay times \cite{Abramavicius:ChemRev:09}: 
\begin{eqnarray}
S^{(3)}_{k_{III}}(\Omega_1,\Omega_2,t_3)=\int_0^\infty \mathrm{d} t_1 \int_0^\infty \mathrm{d} t_2 e^{\imath \Omega_1 t_1+\imath \Omega_2 t_2} S^{(3)}_{k_{III}}(t_1,t_2,t_3).
\end{eqnarray}
For the analysis, the double quantum coherence signal $S^{(3)}_{k_{III}}$ is plotted as a function of the frequencies
$\Omega_1$, $\Omega_2$, cp. Fig. \ref{lioupathways}a):
\begin{eqnarray}
&&S^{(3)}_{k_{III}}(\Omega_1,\Omega_2,t_3)= S^{(3)}_{i}(\Omega_1,\Omega_2,t_3)+S^{(3)}_{ii}(\Omega_1,\Omega_2,t_3) \label{k3first}
\\
 &&S^{(3)}_{i}(\Omega_1,\Omega_2,t_3)\nonumber\\
 &&\quad=\frac{1}{\hbar^3}\sum_{ee'f}\mu_{e'f}\cdot E^{4*}(\omega_{fe'}) 
\mu_{ge'}\cdot E^{3*}(\omega_{e'g})
\nonumber\\ &&\qquad\qquad 
 \mu^*_{ef}\cdot E^{2}(\omega_{fe})
\mu_{ge}^*\cdot  E^{1}(\omega_{eg}) 
 \label{k3einsfirst}
\frac{\mathrm{exp}({-\imath \xi_{fe'} t_3}) }{  (\Omega_2-\xi_{fg})(\Omega_1-\xi_{eg})} 
 \end{eqnarray}
 \begin{eqnarray}
&& S^{(3)}_{ii}(\Omega_1,\Omega_2,t_3)\nonumber\\
&&\quad=-\frac{1}{\hbar^3}\sum_{ee'f}\mu_{ge'}\cdot
E^{4*}(\omega_{e'g})
\mu_{e'f}\cdot E^{3*}(\omega_{fe'})    \nonumber\\ &&\qquad\qquad 
 \mu^*_{ef}\cdot E^{2}(\omega_{fe}) 
\mu_{ge}^*\cdot  E^{1}(\omega_{eg})   \frac{\mathrm{exp}({-\imath \xi_{e'g} t_3})}{  (\Omega_2-\xi_{fg})(\Omega_1-\xi_{eg})}.  \label{k3zweifirst}
\end{eqnarray}

It  exhibits resonances for the ground state-single-exciton transitions $\omega_{eg}$ along the $\Omega_1$ 
axis and the  ground state-two-exciton transition $\omega_{fg}$ \cite{Yang:PhysRevLett:08,Abramavicius:ChemRev:09} along  the $\Omega_2$ axis. Due to the use of the local oscillator, the imaginary and real part of $S^{(3)}_{k_{III}}(\Omega_1,\Omega_2,t_3)$ can be obtained from experimental data \cite{Abramavicius:ChemRev:09}.

\subsection{Localized double quantum coherence signal}
For localized spectroscopy described here, the double quantum coherence signal $S^{(3)}_{k_{III}}$\cite{Yang:PhysRevLett:08,Abramavicius:ChemRev:09} is 
modified by localizing the first pulse at a specific quantum dot $i$,
cf. Fig. \ref{geometry}b).

\noindent For a localized excitation the Hamiltonian Eq. (\ref{licht_matter})
must be modified: 
\begin{eqnarray}
H_{el-L}=\sum_i \mu_{gi}\cdot E(r_i,t) |g \protect\rangle  \protect\langle i|+\sum_{ij} \mu_{gi}\cdot E(r_i,t) |j \protect\rangle_l  \protect\langle ij|_l +H.a.
\end{eqnarray}
 and
 yields 
 \begin{eqnarray}
 H_{el-L}=\sum_{ie} c^{e}_i\mu_{gi}\cdot E(r_i,t) |g \protect\rangle  \protect\langle e|+\sum_{i<j ef} {c^{e}_i}^* \mu_{gi}c^{f}_{ij}\cdot E(r_i,t) |e \protect\rangle  \protect\langle f| +H.a.
 \end{eqnarray}
 for the delocalized states. We see that no delocalized dipole moments are formed, since the effective response depends on the spatial distribution of the electric field.
 
Using far field excitation for pulses $E^2$, $E^3$, the local oscillator $E^4$ for heterodyne detection and a localized excitation for the first pulse $E^1$ at dot $i$ ($E^{1}\rightarrow E^{1}_i$),  the double quantum coherence signal 
$S^{(3)}_{k_{III}}(i,\Omega_1,\Omega_2,t_3)= S^{(3)}_{i}(i,\Omega_1,\Omega_2,t_3)+S^{(3)}_{ii}(i,\Omega_1,\Omega_2,t_3)$ now dependents on the chosen quantum dot $i$ and reads:
\begin{eqnarray}
&&S^{(3)}_{k_{III}}(i,\Omega_1,\Omega_2,t_3)= S^{(3)}_{i}(i,\Omega_1,\Omega_2,t_3)+S^{(3)}_{ii}(i,\Omega_1,\Omega_2,t_3) \label{k3locfirst}
\\
 &&S^{(3)}_{i}(i,\Omega_1,\Omega_2,t_3)\nonumber\\
 &&\quad=\frac{1}{\hbar^3}\sum_{ee'f j}\mu_{e'f}\cdot E^{4*}(\omega_{fe'}) 
\mu_{ge'}\cdot E^{3*}(\omega_{e'g})
\nonumber\\ &&\qquad\qquad 
 \mu^*_{ef}\cdot E^{2}(\omega_{fe})
c^{e*}_{j} 
\mu_{gj}^*\cdot  E^{1}_i(\mathbf{r}_j,\omega_{eg}) 
 \label{k3einslocfirst}
\frac{\mathrm{exp}({-\imath \xi_{fe'} t_3}) }{  (\Omega_2-\xi_{fg})(\Omega_1-\xi_{eg})}, 
 \end{eqnarray}
 \begin{eqnarray}
&& S^{(3)}_{ii}(i,\Omega_1,\Omega_2,t_3)\nonumber\\
&&\quad=-\frac{1}{\hbar^3}\sum_{ee'f j}\mu_{ge'}\cdot
E^{4*}(\omega_{e'g})
\mu_{e'f}\cdot E^{3*}(\omega_{fe'})    \nonumber\\ &&\qquad\qquad 
 \mu^*_{ef}\cdot E^{2}(\omega_{fe}) 
c^{e*}_{j}
\mu_{gj}^*\cdot  E^{1}_i(\mathbf{r}_j,\omega_{eg})   \frac{\mathrm{exp}({-\imath \xi_{e'g} t_3})}{  (\Omega_2-\xi_{fg})(\Omega_1-\xi_{eg})}.  \label{k3zweilocfirst}
\end{eqnarray}
 $\mu_{eg}$/$\mu_{fe}$ are single-exciton/two-exciton to ground state/single-exciton dipoles in the delocalized basis and $\mu_{gi}$ is the dipole moment for the ground state to excited state transition of quantum dot $i$.
$E^1_i(\mathrm{r}_j,\omega_{eg})$ is the first pulse,  predominantly exciting quantum dot $i$, only weakly exciting the other quantum dots with $i\neq j$. We assume ideal localization by taking $E^{1}_i(\mathbf{r}_j,\omega_{eg})\approx \delta_{ij} E^{1}_i(\mathbf{r}_i,\omega_{eg})$.

\subsection{Discussion of the  double quantum coherence signal}
Fig. \ref{SDQClocimag} shows the far-field double quantum coherence signal  ($E^1_i(\mathbf{r}_j,\omega)\approx E^1(\omega)$, cf. Sec. \ref{doublequantumcoherencesignal}) absolute Fig. 4a) and imaginary value b):
The frequency of the single exciton to ground state coherence can be seen on the $\Omega_1$ axis and of the 
two exciton to ground state coherence on the $\Omega_2$ axis. 
Clearly, for the far field excitation in Fig \ref{SDQClocimag}a) and b) we see  resonances connecting 
to coherence of several states $e$ and  $f$. 
If we select a frequency $\Omega_1=\omega_{e_i g}$,
we see along the $\Omega_2$ axis, which specific two exciton states are connected via dipole moments
to the single exciton state $e_i$ and vice versa.
A comparison of the dipole moments connected to  two different peaks
works only roughly, since two Liouville paths interfere  
and the degree of destructive interference is different for every peak.

 A dominant peak ($A$) in the absolut value spectrum (Fig. \ref{standardDQC}a)) is connected to $e_2$ and $f_2$, a second strong peak is connected to $e_2$ and $f_1$   and some further peaks with smaller oscillator strength 
can be seen at a lower single-exciton energy ($e_3$ and $f_2$,$e_1$ and $f_1$). 
$e_1$ and $e_2$ are well resolved,  $e_3$ shows up as a spectral shoulder. This  shows that the system has three single-exciton and three two-exciton states. 

Fig. \ref{SDQClocimag}c) shows the  signal  with the first  pulse  localized  at either quantum dot  $1$, $2$ or  $3$. 
The localization of the first pulse gives information about the single exciton states contributing to the ground state-single exciton transition occuring during the first pulse.
Localization at quantum dot 1 shows that
 all resonances connected to the delocalized exciton state $e_1$ disappear. 
 Overall, this shows, that quantum dot 1 only contributes  strongly to the formation of single exciton state $e_2$ and $e_3$, but not to the build up of $e_1$. Similar information is obtained for excitation of quantum dots 2 and  3 (see other Figs. \ref{SDQClocimag}c)).
E.g. the exciton state $e_2$  is formed by quantum dot 1 and 2. 
Another interesting feature is the peak connecting $e_3$ and $f_1$.
This peak is only visible at the localized spectrum at QD 2 and QD 3 and not in the far field spectrum. This is caused by the fact, that $e_3$ is an antisymmetric delocalized state between QD 2 and 3, seen by the opposite sign of the peak in the QD 2 and QD 3 spectrum. For far-field excitation, these two antiparallel dipole interfere destructively, so that the resonance is not observed.

{\it We next use the localized double quantum coherence to  extract the wavefunction coefficients $c^{e}_{i}$ and therefore all quantum dot interactions. }

\section{Extracting the single exciton wavefunction} \label{extract_wavefunc}
All ingredients are now available to extract the single exciton wavefunction.
We start from the localized signal in Eq. (\ref{k3locfirst}-\ref{k3zweilocfirst}) and see that the sum over $e$ and $j$ prevents us to extract a particular coefficient $c^e_i$. 
Assuming ideal localization of the first pulse at a particular quantum dot $i$ ($E^{1}_i(\mathbf{r}_j,\omega_{eg})\approx \delta_{ij} E^{1}_i(\mathbf{r}_i,\omega_{eg})$) removes the sum over $j$ in Eq. (\ref{k3locfirst}-\ref{k3zweilocfirst}). Of course, any deviation from ideal localization will  result in an error in the measurement of the coefficients (see below).

For removing the sum over $e$ and selecting a particular single exciton state $e$, we choose the frequencies $\Omega_1=\Omega_1^e$ and $\Omega_2=\Omega_2^e$ in a way, that only a specific peak  caused by single-exciton  to ground state $\omega_{eg}$ and two-exciton to ground state coherences $\omega_{fg}$ connected to $e$ contributes, as suggested by the denominators in Eq. (\ref{k3locfirst}-\ref{k3zweilocfirst}). Again, if peaks for different single exciton states overlap, errors are introduced to the reconstruction. (However two dimensional spectroscopy has less spectral overlap than one dimensional spectroscopy, since the peaks are separated by an additional degree of freedom: the additional frequency axis.)
This yields:
\begin{eqnarray}
&&S^{(3)}_{k_{III}}(i,\Omega_1^e,\Omega_2^e,t_3)= S^{(3)}_{i}(i,\Omega_1^e,\Omega_2^e,t_3)+S^{(3)}_{ii}(i,\Omega_1^e,\Omega_2^e,t_3) \label{k3locfirst_extracted}
\\
 &&S^{(3)}_{i}(i,\Omega_1^e,\Omega_2^e ,t_3)\nonumber\\
 &&\quad\approx\frac{1}{\hbar^3}\sum_{e'f }\mu_{e'f}\cdot E^{4*}(\omega_{fe'}) 
\mu_{ge'}\cdot E^{3*}(\omega_{e'g})
\nonumber\\ &&\qquad\qquad 
 \mu^*_{ef}\cdot E^{2}(\omega_{fe})
c^{e*}_{i} 
\mu_{gi}^*\cdot  E^{1}_i(\mathbf{r}_i,\omega_{eg}) 
 \label{k3einslocfirst_extracted}
\frac{\mathrm{exp}({-\imath \xi_{fe'} t_3}) }{  (\Omega_2-\xi_{fg})(\Omega_1-\xi_{eg})} 
 \end{eqnarray}
 \begin{eqnarray}
&& S^{(3)}_{ii}(i,\Omega_1,\Omega_2,t_3)\nonumber\\
&&\quad\approx-\frac{1}{\hbar^3}\sum_{e'f }\mu_{ge'}\cdot
E^{4*}(\omega_{e'g})
\mu_{e'f}\cdot E^{3*}(\omega_{fe'})    \nonumber\\ &&\qquad\qquad 
 \mu^*_{ef}\cdot E^{2}(\omega_{fe}) 
c^{e*}_{i}
\mu_{gi}^*\cdot  E^{1}_i(\mathbf{r}_i,\omega_{eg})   \frac{\mathrm{exp}({-\imath \xi_{e'g} t_3})}{  (\Omega_2-\xi_{fg})(\Omega_1-\xi_{eg})}.  \label{k3zweilocfirst_extracted}
\end{eqnarray}
We see, that here the double quantum coherence signal  is proportional to $c^{e*}_{i} \mu_{gi}^*\cdot E_i^1(\mathbf{r}_i,\omega_{eg})$,i.e. to the strength $c_i^{e*}$ the $i$-th quantum dot contributes to the delocalized wave function.
This fact is used to develop a  scheme to extract the coefficients $c_i^e$ from measured data:\\

\noindent As {\bf input information}  the dipole moment $\mu_{gi}$ 
of the individual uncoupled quantum dots are required, the dipole moments can be measured or calculated.
\\
As {\bf measurement,} carry out the localized double quantum coherence signal $S^{(3)}_{k_{III}}(i,\Omega_1,\Omega_2,t_3)$, for a localization on all quantum dots $i$. If the field strength and polarisation direction is different for localization at different quantum dots, we need to obtain the electric field along the local dipole $\mu_{gi}^*\cdot E_i^1(\mathbf{r}_i, \omega_{eg})$.

\noindent Now, we select the  excitonic state $e\equiv e_{\alpha}$, whose coefficients $c^{e_{\alpha}}_i$ should be extracted.
We determine the  the frequencies $\Omega_1\approx \omega_{e_{\alpha}g}$, 
$\Omega_2\approx\omega_{f_{\beta}g}$ showing a strong correlation to $e_{\alpha}$ 
using the double quantum coherence signal without spatial localization.

 \noindent Now in the {\bf postprocessing} of the data, 
 we use that $c^{e_{\alpha}*}_{i}\propto S^{(3)}_{k_{III}}/(\mu_{gi}^*\cdot E_i^1(\mathbf{r}_i, \omega_{eg}))$ at the positions $\Omega_1\approx \omega_{e_{\alpha}g}$,  
 $\Omega_2\approx\omega_{f_{\beta}g}$ (Eq. (\ref{k3locfirst_extracted}-\ref{k3zweilocfirst_extracted})).
$c^{e_{\alpha}*}_{i}$ can now be determined up to an proportionality factor $A$:  
$c^{e_{\alpha}*}_{i} A = S^{(3)}_{k_{III}}/(\mu_{gi}^*\cdot E_i^1(\mathbf{r}_i, \omega_{eg}))$ for every quantum dot $i$, using the same frequencies $\Omega_1$, $\Omega_2$. 
 Since the wavefunction is normalized, $|A|^2=\sum_i|A c^{e_\alpha *}_{i}|^2$ holds.  We thus get
 $A$ up to a global phase and set $A=|A|$.
 We obtain $c^{e*}_{i}
  =S^{(3)}_{k_{III}}(i,\Omega_1,\Omega_2,t_3)/(\mu_{gi}^* \cdot E_i^1(\mathbf{r}_i, \omega_{eg})) A)$.
This gives the delocalized wavefunction $|e_\alpha\rangle=\sum_i c^{e_\alpha}_i |i\rangle$.\\
Note, that these steps constitutes a  quantum state tomography.
 The local basis is uniquely determined  up to an arbitrary phase for every quantum dot: the expansion coefficient $c^e_i$ depend on that choice.

\begin{figure}[tbh]
\center
\includegraphics[width=8 cm]{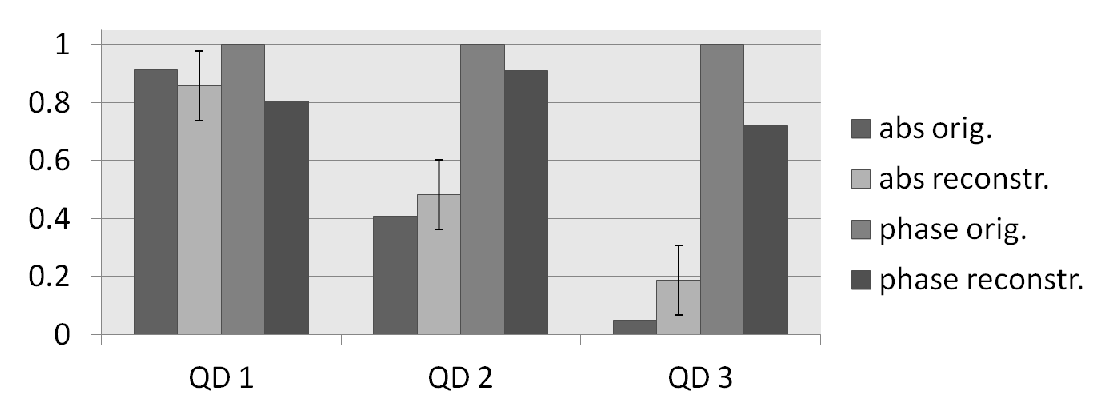} 
\caption{  Original and reconstructed coefficients of  single-exciton wave function  $e_2$: 
Phase in multiples of $2\pi$.  Error of absolute values determined by localization.
} 
\label{wavefunctionreconstr}
\end{figure}

To demonstrate the success of the tomography, we compare in Fig. \ref{wavefunctionreconstr}
the elements of the   reconstructed wavefunction  for the strongest contribution, i.e. state $e_2$ (marked with A in Fig. \ref{standardDQC}a)),
to the original  wave function  resulting from the input parameters in the Hamiltonian. 
The agreement for both the amplitude and the relative phase is quite good. 
The difference results from  a non-perfect localization $E^{1}_i(\mathbf{r}_j,\omega_{eg})\neq \delta_{ij} E^{1}_i(\mathbf{r}_i,\omega_{eg})$ resulting from realistic
Maxwell simulation from section \ref{localized_excitation}. 
This error is marked by the error bars in Fig.  \ref{wavefunctionreconstr}. 
 It is caused by a weak excitation of quantum dots,  which a ideally localized pulse should not excite. 
 Such a non ideal excitation leads to a cross talk between the coefficients. The error bars are  estimated to be smaller than: $\Delta_{c^e_i}=\sum_{j\neq i} |E_i(\mathrm{r}_j)|/|E_i(\mathrm{r}_i)|$.
 
 Note, that in general,
 the procedure works also for other methods than heterodyne detection in far field, including a localized detection of polarisation or fluorescence, as long as the detection is the same for a localization of the first pulse at different quantum dots.
The only limitation is, that
 the phase of the coefficients can only be detected with methods, that can measure complex signals.
For other types of detection like homodyne detection, we can also extract the absolute value of the coefficients, but not their phase.

\section{Filtering coherent spectra}
As additional useful application, we show that strong, undesired resonances can be selectively suppressed from coherent spectra.
This can be advantageous while investigating
 weak resonances, that are masked by other strong resonances:
 Often, it is not clear, whether weak resonances constitute
 a vibrational side peak connected to a dominanting strong excitonic peak
 or a different, much weaker excitonic resonance.
This can also  be  solved by selectively removing excitonic resonances from measured spectra,
applying a filter algorithm.\\
As {\bf input information} for the filter algorithm, we have
to determine expansion coefficients $c^{e_{\alpha}}_{i}$ for the specific state $e_{\alpha}$ for all quantum dots $i$,
whose contributions we want to filter out.
Additionally, we need all dipole moments $\mu_{gi}$ of the individual nanostructure and also the electric field along the local dipole $\mu_{gi}^*\cdot E_i^1(\mathbf{r}_i, \omega_{eg})$.\\
As {\bf measurement} we record a localized version of the spectrum to be filtered. The localized pulse should excite a ground state to single exciton transition for all quantum dot positions. For the double quantum coherence, this will be 
 the signal $S_{k_{III}}(i,\Omega_1,\Omega_2,t_3)$ for every quantum dot $i$.\\
For {\bf postprocessing} we discuss the expression 
\begin{eqnarray}
&& S_{k_{III},w/o e_{\alpha}}(\Omega_1,\Omega_2,t_3)\nonumber\\
&& \qquad =S_{k_{III}}(\Omega_1,\Omega_2,t_3)-\sum_{i} {c_i^{e_{\alpha}}}^*
\mu_{ig}^* \cdot E^1(\omega_{eg})
f_{e_{\alpha}}(\Omega_1,\Omega_2,t_3) \nonumber\\
&&\qquad=\sum_{e'\neq e_{\alpha}, i} {c_i^{e'}}^*
\mu_{ig}^* \cdot E_i^1(\omega_{eg})
f_{e'}(\Omega_1,\Omega_2,t_3) \label{filtered_spectrum}\\
&& f_{e_{\alpha}}(\Omega_1,\Omega_2,t_3)=\sum_i c^{e_{\alpha}}_i
S_{k_{III}}(i,\Omega_1,\Omega_2,t_3)/(\mu_{ig}^*  \cdot E_i^1(\mathbf{r}_i,\omega_{eg})) \label{fdefinition}
\end{eqnarray}
which gives a spectrum, where all contribution of $e_{\alpha}$ during the first pulse are filtered out.
\footnote{The idea behind the filtering algorithm is, that we can write the localized spectrum  in terms of individual contributions $f_{e'}$ caused by resonances for different single-excitons $e'$, where $f_{e'}$ are calculated from measured spectra (cf. Eq. (\ref{fdefinition})):
\begin{eqnarray}
 S_{k_{III}}(i,\Omega_1,\Omega_2,t_3)&=&\sum_{e'} {c_i^{e'}}^* \mu_{ig}^*\cdot E_i^1(\mathbf{r}_i,\omega_{eg})
f_{e'}(\Omega_1,\Omega_2,t_3)
\end{eqnarray}
 Multipling the equation with  $c^{e}_i$ and summing over $i$ yields a scalar product and
 we get $f_{e'}$ defined using the localized spectra (Eq. (\ref{fdefinition})). 

The far field double quantum coherence spectrum can also be calculated using $f_{e'}$:
\begin{eqnarray}
 S_{k_{III}}(\Omega_1,\Omega_2,t_3)&=&\sum_{e'} {c_i^{e'}}^* \mu_{ig}^*\cdot E^1(\omega_{eg})
f_{e'}(\Omega_1,\Omega_2,t_3).
\end{eqnarray}
It is expressed using summands for every contributing single exciton $e'$, the summand of the exciton to be filtered can be  substracted. This is possible, since every summand can be calculated using $f_{e'}$, which can be calculated from the experimental data of the  localized spectrum, if  the expansion coefficients of the single exciton wavefunction of $e'$ are known.}

The  single-exciton peak $e_2$ ($\alpha=2$)  dominating the spectrum in Fig. \ref{standardDQC} a) is filtered out
in Fig. \ref{standardDQC}d). 
This spectrum reveals now information  about states initially covered by the dominant contribution of $e_2$.
The procedure can be applied iteratively, using the filtered spectra  for obtaining the other excitonic states.
This can enhance the reconstruction of the exciton states.

The filtering method can also be applied to other spectroscopic signals as long as a phase sensible detection is used and a localized signal, whose contributions are proportional to the single exciton expansion coefficients, can be measured.

\section{Conclusion and outlook}
The presented quantum state tomography for the extraction of the delocalized single exciton wave function coefficients, can  also be applied to other impulsive two dimensional spectra.
The single-exciton to two-exciton transition in double quantum coherence using the localization of the second pulse also also can be used to extract the  two exciton coefficients. However since this problem is more complex, it will be subject to future work.

In conclusion, our simulations demonstrate a quantum state tomography  that
can be used to reconstruct individual wave functions of  coupled emitters acting only collectively in the far field.
In addition, localized excitations are useful  to remove unwanted strong resonances  to uncover weak or hidden excitonic resonances. All of these features are not accessible in standard far field spectroscopy.
  Similar configurations can be alternatively achieved by applying four pulses and using phase cycling to detect a desired component\cite{Brinks:Nature:10}  e.g. with phase $\varphi=\varphi_1+\varphi_2-\varphi_3-\varphi_4$.
We therefore believe that the proposed quantum state tomography opens a new path for the detection  of many body interactions on the nanoscale.
The proposed protocol is more general as presented here, since fluorescence can also be used  rather than heterodyne detection of optical fields \cite{Brinks:Nature:10}.

\begin{acknowledgments}
We gratefully acknowledge support from the Deutsche Forschungsgemeinschaft (DFG) through SPP 1391 (M.R., Fe.S., Fr.S.), GRK 1558 (M.S.), SFB 951 (A.K.).
M.R. also acknowledges support from the Alexander von Humboldt Foundation through the Feodor-Lynen program.
S.M. gratefully acknowledges the support of NSF grant CHE-1058791,
DARPA BAA-10-40 QUBE, and  the Chemical Sciences, Geosciences and
Biosciences Division, Office of Basic Energy Sciences, Office of Science, (U.S.) Department of Energy (DOE). 
We also thank Jens Förstner and Torsten Meier, Paderborn for very valuable discussion about the application of the genetic algorithm for the localized fields.
\end{acknowledgments}

\end{document}